%% file: manuscript.tex
\documentclass[10pt]{article}

\usepackage{amsmath,amssymb,graphicx,cite,color} 

\usepackage[labelfont=bf,labelsep=period,justification=raggedright]{caption}

\makeatletter
\renewcommand{\@biblabel}[1]{\quad#1.}
\makeatother

\date{}

\newcommand{\eref}[1]{(\ref{eqn:#1})}
\newcommand{\fref}[1]{Fig.~\ref{fig:#1}}
\newcommand{\sfref}[1]{Suppl.~Fig.~\ref{fig:#1}}
\newcommand{\tref}[1]{Table~\ref{tab:#1}}

\begin{document}

% Title must be 150 characters or less
\begin{flushleft}
{\Large
\textbf{Mean-field dynamics of tumor growth and control using low-impact chemoprevention}
}
% Insert Author names, affiliations and corresponding author email.
\\
Andrei R. Akhmetzhanov$^{1,2}$, 
Michael E. Hochberg$^{1,3,4,5,\ast}$
\\
\bf{1} Institute of Evolutionary Sciences of Montpellier - UMR 5554, University of Montpellier II, CC065, Place Eug\'ene Bataillon, 34095 Montpellier Cedex 5, France
\\
\bf{2} Theoretical Biology Lab, Dept. of Biology, McMaster University, Hamilton, Ontario L8S4K1, Canada
\\
\bf{3} Santa Fe Institute, Santa Fe, NM 87501, USA
\\
\bf{4} Wissenschaftskolleg zu Berlin, Wallotst. 19, 14193 Berlin, Germany
\\
\bf{5} Kavli Institute for Theoretical Physics, University of California, Santa Barbara, CA 93106-4030, USA
\\
$\ast$ E-mail: mhochber@univ-montp2.fr
\end{flushleft}

 \section*{Abstract} 

Cancer poses danger because of its unregulated growth, development of resistant subclones, and metastatic spread to vital organs. Although the major transitions in cancer development are increasingly well understood, we lack quantitative theory for how chemoprevention is predicted to affect survival. We employ master equations and probability generating functions, the latter well known in statistical physics, to derive the dynamics of tumor growth as a mean-field approximation. We also study numerically the associated stochastic birth-death process. Our findings predict exponential tumor growth when a cancer is in its early stages of development and hyper-exponential growth thereafter. Numerical simulations are in general agreement with our analytical approach. We evaluate how constant, low impact treatments affect both neoplastic growth and the frequency of chemoresistant clones. We show that therapeutic outcomes are highly predictable for treatments starting either sufficiently early or late in terms of initial tumor size and the initial number of chemoresistant cells, whereas stochastic dynamics dominate therapies starting at intermediate neoplasm sizes, with high outcome sensitivity both in terms of tumor control and the emergence of resistant subclones. The outcome of chemoprevention can be understood in terms of both minimal physiological impacts resulting in long-term control and either preventing or slowing the emergence of resistant subclones. We argue that our model and results can also be applied to the management of early, clinically detected cancers after tumor excision. \section*{Author summary} 

One of the principal risks of aggressive chemotherapy is the selection of cells that are at the origin of relapse and are refractory to subsequent treatments. Alternative approaches based on management models usually target clinically detected tumours or residual cancers. These therapies carry the risk of not being able to control fast growing subclones or resistant lineages. We develop a mean-field approach to evaluate low impact chemoprevention. Effective management slows or prevents evolution though the incorporation of fitness-enhancing driver mutations, and the emergence of chemoresistance. Dynamics are highly predictable for sufficiently small or large initial tumour sizes, and increasingly stochastic for intermediate-sized tumours. Based on empirical parameter estimates, we predict that the optimal daily levels of reduction in tumour growth for sufficiently small neoplasms are between \(0.1\%\) and \(0.2\%\). This corresponds to reducing the net growth rate of the existing tumour to below zero (\(0.1\%\) growth reduction), but not so much so as to select for subsequent driver mutations (\(0.2\%\) growth reduction). Satisficing based on chemoprevention offers an alternative approach for people at high risks of life-threatening cancers. \section{Introduction}

Mathematical models play an important role in describing and analyzing the complex process of carcinogenesis. Natural selection for increases in tumor cell population growth rate can be represented as the net effect of increased fission rates and/or decreased apoptosis (e.g., \cite{WodaKoma07}). Relatively rare driver mutations confer such a net growth advantage, whereas numerically dominant passenger mutations with initially neutral or mildly deleterious effects \cite{BoziAnta10,MaruAlme12,McFaKoro13} can only initially grow in frequency due to genetic hitchhiking. Amongst the many passengers in a growing tumor, some can contribute to cell chemoresistance, and a sufficiently large tumor will contain different clones that, taken as a group, can resist most, if not all, possible chemotherapies (see \cite{MichHugh05} for resistance to imatinib). Chemotherapeutic remission followed by relapse suggests that these resistant cells are often at low frequencies prior to therapy, either due to genetic drift or costs associated with resistance. Resistant phenotypes subsequently increase in frequency during chemotherapy, and through competitive release, they may incorporate one or more additional drivers, resulting in accelerated growth compared to the original tumor \cite{HuijBell13}.

Previous mathematical studies have considered alternatives to attempting to minimize or eradicate clinically diagnosed cancers with maximum tolerated doses (MTD) of chemotherapeutic drugs. This body of work indicates that MTD is particularly prone to select for chemoresistance (e.g., \cite{FooMich09, FooMich10, LorzLore13}), and empirical studies support this basic prediction \cite{TurkZejn10}. Numerous alternatives to the goal of cancer minimization/eradication have been investigated (e.g., \cite{FooMich09, KomaWoda05, BoziReit13, GateBrow09, MaleReid04}). For example, Komarova and Wodarz \cite{KomaWoda05} considered how the use of one or multiple drugs could prevent the emergence or curb the growth of chemoresistance. They showed that the evolutionary rate and associated emergence of a diversity of chemoresistant lineages is a major determinant in the success or failure of multiple drugs versus a single one. Foo and Michor \cite{FooMich09} evaluated how different dosing schedules of a single drug could be used to slow the emergence of resistance given toxicity constraints. One of their main conclusions is that drugs slowing the generation of chemoresistant mutants and subsequent evolution are more likely to be successful than those only increasing cell death rates.

These and other computational approaches have yet to consider the use of chemoprevention to reduce cancer-associated morbidity and mortality. Prevention, more generally, encompasses life-style changes, interventions or therapies in the absence of detectable invasive carcinoma (e.g., \cite{EtziUrba03, LippLee06, WillHeym09, HochThom13}). In depth consideration of preventive measures and their likely impact on individual risk and epidemiological trends is important given the virtual certitude that all people have pre-cancerous lesions, some of which may transform into invasive carcinoma \cite{BissHine11, Grea14}, and concerns as to whether technological advances will continue to make significant headway in treating clinically detected cancers \cite{VogePapa13, GillFlow12}. 

Here we model how chemoprevention affects tumor progression and the emergence of chemoresistant lineages. Previous study has considered the effects of deterministic and stochastic processes on tumor growth and the acquisition of chemoresistance \cite{KomaWoda05,BoziAnta10,ReitBozi13}. We consider both processes through exact solutions and numerical simulations of master equations, using the mean field approach. A mean field approach assumes a large initial number of cells \cite{KrapRedn10} and averages any effects of stochasticity, so that an intermediate state of the system is described by a set of ordinary differential equations (i.e., master equations; \cite{Gard04}). Solutions to these are complex even in the absence of the explicit consideration of both drivers and passengers \cite{AntaKrap11}. Our approach \cite{BaakWagn01,SaakHuC06} follows the dynamics of the relative frequencies of subclones, composed of identical cells, instead of the fate of individual cells. We derive the dynamics for the expected total number of cells within a tumor at any given time. 

We show that the expected mean tumor size can be substantially different from the median, since the former is highly influenced by outliers due to tumors of extremely large size. We then consider constant chemopreventive treatments, starting at a given tumor size and number of chemoresistant mutations. We find that treatment outcome can be highly sensitive to initial conditions. Not surprisingly, initially small tumors are more likely to be controlled than larger tumors employing low dose therapies, whereas large tumors follow deterministic growth and are both difficult to control in overall size and in the emergence of resistance. In contrast, there is a range of intermediate size tumors, where stochastic dynamics become significant, and clinical outcome is highly sensitive to the commencement time (i.e., initial tumor size) of treatment regimes. \section{Results}

\begin{table}[!bth]\input{table1.tex}\caption{Canonical parameter values used in this study.}\label{tab:1}\end{table}

We study a low intensity, constant treatment regime that starts at time \(t=0\). First, we study mean-field dynamics by considering the distribution \(H_t(x)\) of tumor sizes \(x\) at time moments \(t\), and examine effects on the mean \(n(t)=\langle H_t(x)\rangle\). 

Using the master equations, we derive an analytical expression for the dynamics \(n(t)\). Namely, we use \eref{2.2.3}-\eref{2.2.4} and \eref{2.3.4}-\eref{2.3.5} (see Methods section) to obtain the dynamics of the expected tumor size
\begin{equation}\label{eqn:3.1.1} n(t) = n(0)\left((1-\kappa)\left(1+\frac{v}2\frac{e^{(\sigma-c)t}-1}{\sigma-c}\right)+\kappa e^{(\sigma-c)t}\right)\exp\left[(s-\sigma)t + N\ln\!\left(1+\frac{u}{2N}\frac{e^{st}-1}{s}\right)\right]\,,    \end{equation} 
and the frequency of resistant cells within a tumor 
\begin{equation}\label{eqn:3.1.2} \frac{n_{res}(t)}{n(t)} = \frac{(1-\kappa)\frac{v}{2}\frac{e^{(\sigma-c)t}-1}{\sigma-c}+\kappa e^{(\sigma-c)t}}{(1-\kappa)(1+\frac{v}{2}\frac{e^{(\sigma-c)t}-1}{\sigma-c})+\kappa e^{(\sigma-c)t}}\,.    \end{equation}
Here, time \(t\) is normalized so that any event occurs at rate equal to unity, or \(t=\tau/T\). In the following, we use the variable \(t\) as shorthand for \(t=\tau/T\).

\fref{3.1.1}  (\textbf{A} and \textbf{a}) shows the excellent correspondence between numerical experiments and analytical results for \(\sigma\) on the order of \(s\).

\begin{figure}[!tbh]\centering{\includegraphics[scale=0.6]{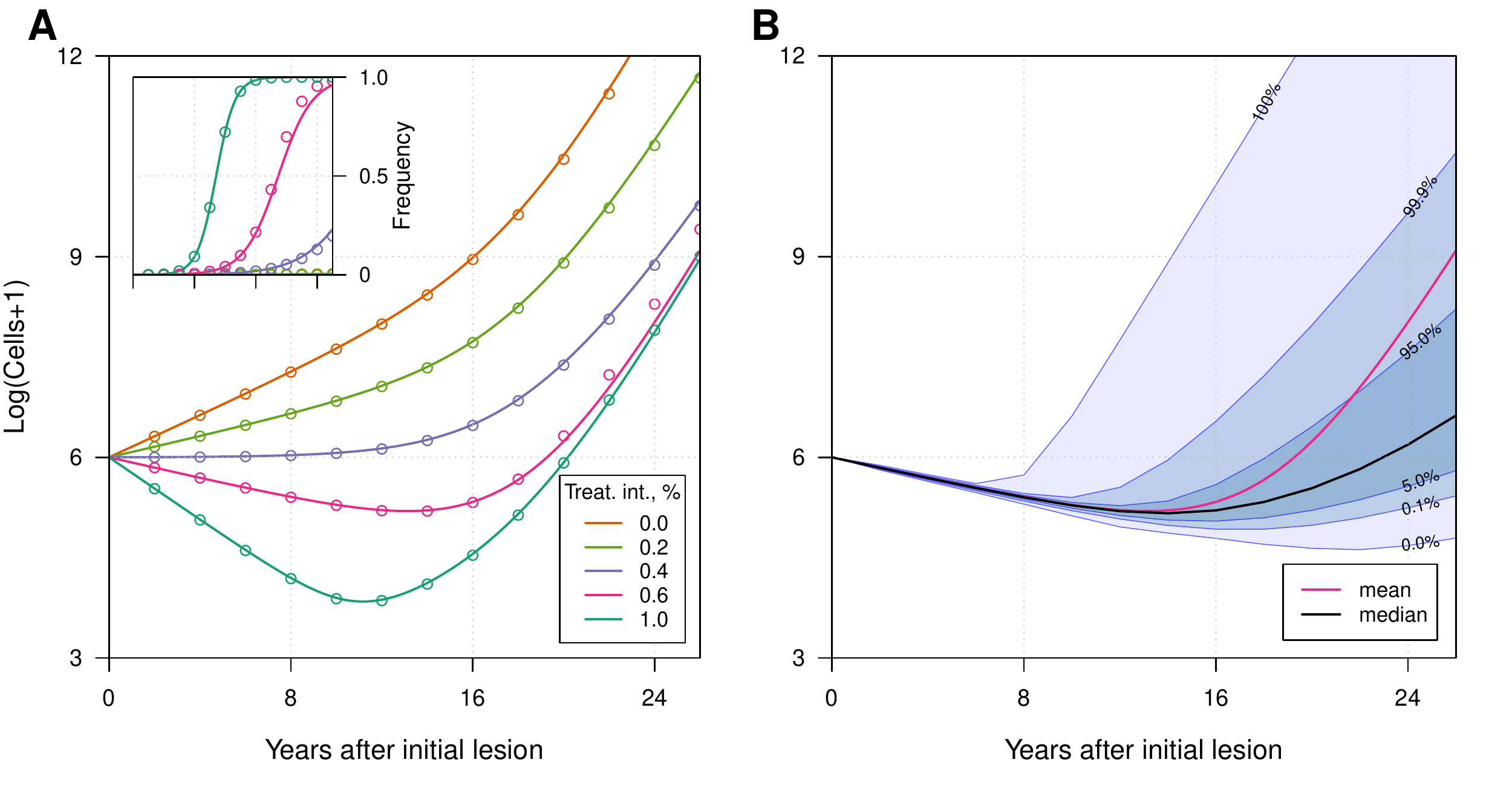}\quad}\caption{\textbf{Mean field dynamics concord with numerical simulations.} (\textbf{A}) Effect of treatment level and observation time on mean tumor size. (\textbf{Inset}) Mean frequency of resistant cells within tumors corresponding to three of the cases in \textbf{A}. Lines are analytically computed mean-field trajectories, while dots are numerical simulations (see Methods section for details). (\textbf{B}) Dynamics of mean and median tumor size, and percentiles around the mean (shaded areas), assuming a fixed constant treatment of \(0.6\%\). Treatments start at \(t=0\), the maximal number of additionally accumulated drivers is 3. See \tref{1} for other parameter values.}\label{fig:3.1.1}\end{figure}

Equation \eref{3.1.1} is simplified for two limiting cases. In the early stages of tumor growth, the value \(n(t)\) changes according to a hyper-exponential law: 
\[ n(t) \approx n(0)\left((1-\kappa)\left(1+\frac{v}2\frac{e^{(\sigma-c)t}-1}{\sigma-c}\right)+\kappa e^{(\sigma-c)t}\right)\exp\!\left[(s-\sigma)t+\frac{u}2\frac{e^{st}-1}s\right]\,. \]
while at later stages the most aggressive subclone persists, being sensitive if \(\sigma < c\) (\(n(t)\propto e^{s(N+1)t}\)) and resistant otherwise (\(n(t)\propto e^{s(N+1)t-c}\)).

A more detailed study of the distribution \(H_t(x)\) reveals that the mean \(n(t)\) diverges importantly from median behavior in the majority of cases, since the former is strongly influenced by outliers. \fref{3.1.1}(\textbf{B}) and \sfref{3.1.S1} illustrate examples where the mean trajectory deviates from the median, and exceeds the 95\% confidence interval at approximately 22 years into the simulation. 

To see how alternative formulations affect the results, we investigate \(H_x(t)\), which is the distribution of times \(t\) when a tumor reaches a given threshold size \(x\). We assume \(x=M=10^9\) cells (i.e., the lower boundary for clinical detection of a tumor - approximately \(1 cm^3\) in volume). Note that the results below can be generalized for other values of \(M\). 

Based on extensive numerical experiments, we find that the means of the distributions \(\langle H_t(x)\rangle\) and \(\langle H_x(t)\rangle\) (as well as other characteristics such as the mode and the median) are the same only in case of no treatment (\(\sigma=0\)). The reason for these apparent discrepancies is that a tumor has two distinct subpopulations (sensitive and resistant), meaning that the distributions \(H_t(x)\) and \(H_x(t)\) are bimodal. Trivially, this does not occur for \(\sigma=0\), since the resistant part is negligible (at a mutation-selection balance).

We perform three sets of numerical experiments to study how variation in any one of the following parameters--the selective advantage \(s\), the cost of resistance \(c\), or the initial number of cells \(n(0)\)--influences the properties of the distribution \(H_M(t)\). 

Variation in the selective advantage \(s\), with \(c=0.1\%\) and \(n(0)=10^6\) cells being fixed, leads to \fref{3.1.3}. We see that tumor growth is mainly driven by its non-resistant part for relatively low impact treatments \(\sigma < 2s\). The tumor changes from being mainly non-resistant to resistant at \(\sigma\approx2s\), which is reflected by the emergence of an inflection point in the trajectory of the median (indicated by \(C\) in \fref{3.1.3}). Notice that the detection times are also most variable at \(\sigma\approx2s\). The median changes smoothly at high treatment levels (\(\sigma>2s\)), tending to a horizontal asymptote. This is explained by the fact that the sensitive part is heavily suppressed at high treatment levels, meaning that the dynamics are strongly influenced by an actual time point when the resistance mutation occurs. 

The inflection point at \(\sigma\approx2s\) is due to the accumulation of additional drivers within tumors and associated increases the likelihood that the tumor eventually resists treatment if no resistant cells were initially present. Since the initial population consists of \(10^6\) cells, in the absence of treatment, a new cell with one additional driver and associated fitness \((2s-\sigma)\) will appear very rapidly. Such a tumor can only be suppressed only if we apply the treatment with \(\sigma>2s\). This is supported by additional numerical experiments, where we vary the maximal number of allowed driver mutations \(N\), see \fref{3.1.9}(\textbf{A}). We see that the inflection point \(\sigma\approx2s\) disappears when \(N=0\). Similarly, we may expect inflection points around \(\sigma=3s\), \(4s\) and so on, which is shown in \sfref{3.1.S2} or in Suppl. Video S3  and where the resistant mutation is knocked out. In contrast, when the resistant mutation is present, an appearance of many inflection points is blurred by higher growth of the resistant part of a tumor: only two humps are noticed in \fref{3.1.3} for the largest value of cost of resistance \(c=0.4\%\) and one hump for smaller values of \(c\).

\begin{figure}[!tbh]\centering{\includegraphics[scale=0.51]{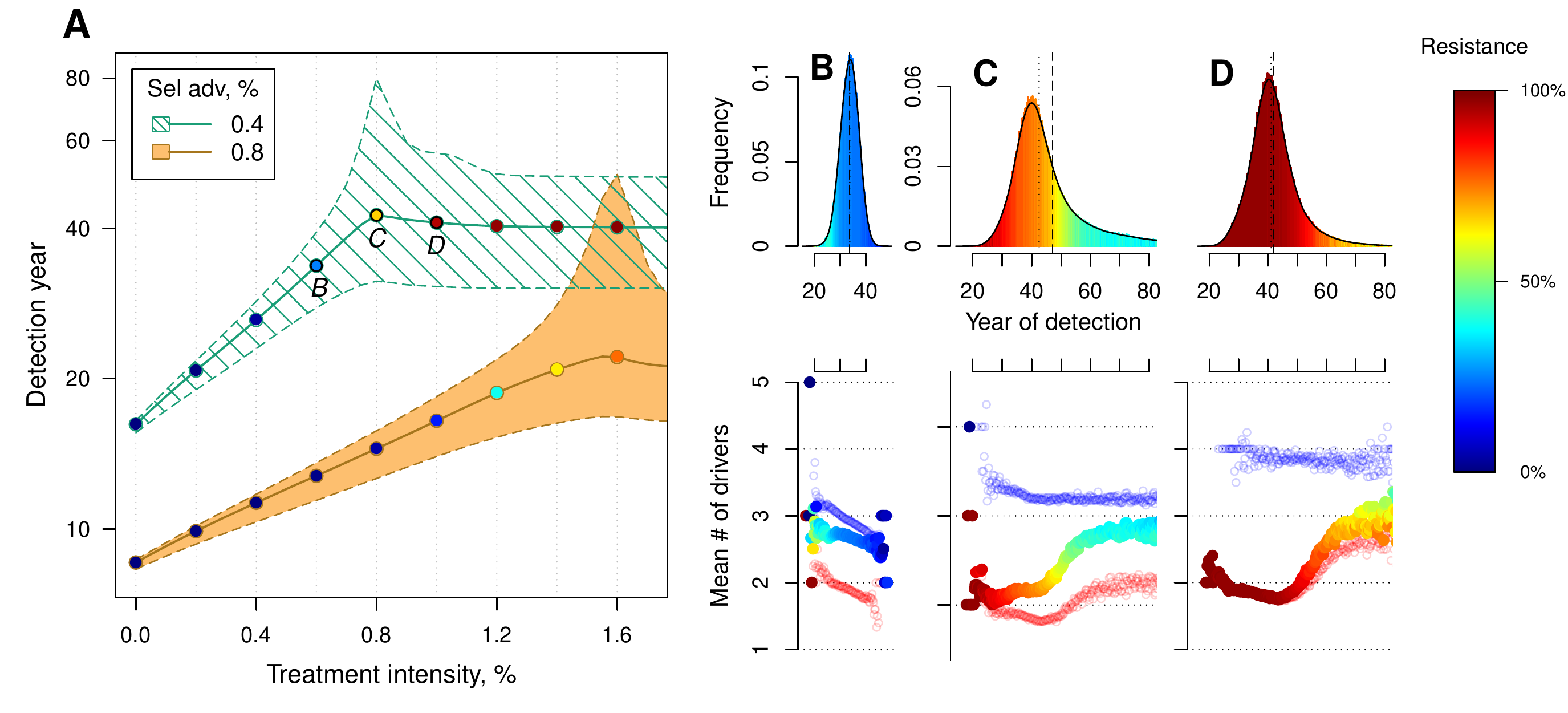}\quad}\caption{\textbf{Treatment level affects both detection time and frequency of resistance.} (\textbf{A}) The median and 95\% confidence intervals (shaded or hatched areas) of detection times for \(0.4\%\) and \(0.8\%\) selective advantages (see legend). (\textbf{B}, \textbf{C} and \textbf{D}) Three samples of the distribution of detection times for corresponding points \emph{B}, \emph{C} and \emph{D}, shown in \textbf{A}. Dashed black line is the mean and the dotted line is the median. Bottom panel shows the mean number of additionally accumulated drivers within tumors over periods of 3 months. Light red points correspond to tumors with a majority of resistant cells; light blue points are for tumors with a majority of non-resistant cells. Color-code indicates the level of resistance in detected tumors over 3 month intervals. No pre-resistance is assumed. Other parameter values are as in \tref{1}. Note that the detection time in \textbf{A} is log-transformed.}\label{fig:3.1.3}\end{figure}

\begin{figure}[!tbh]\centering{\includegraphics[scale=0.51]{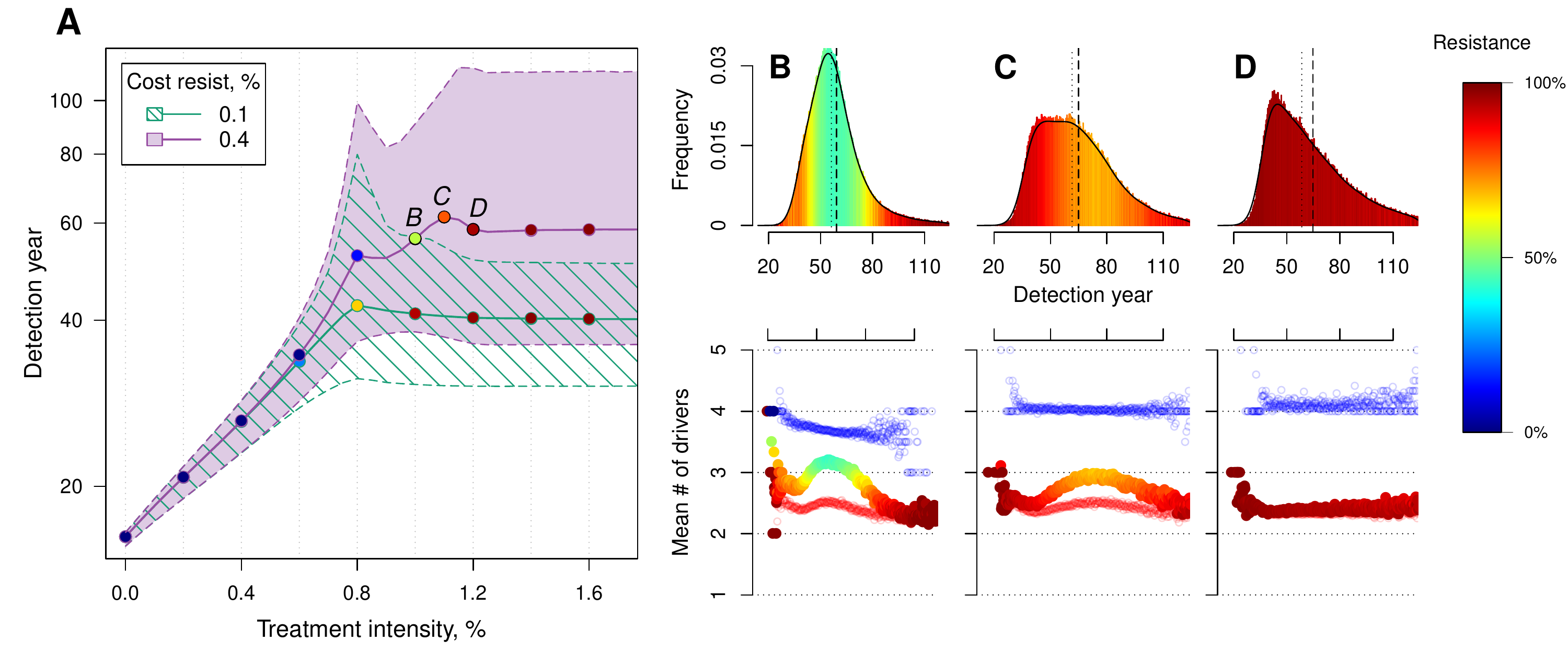}\quad}\caption{\textbf{Higher costs of resistance lead to slower, but more variable growth.} The median and 95\% confidence intervals for detection times when the cost of resistance is varied. The selective advantage is fixed at \(0.4\%\). Other conditions as for \fref{3.1.3}.}\label{fig:3.1.4}\end{figure}

Counterintuitively, if the cost of resistance is low to moderate then early-detected tumors are more likely to be resistant under constant treatments than those detected at later times (\textbf{B}, \textbf{C} and \textbf{D} in \fref{3.1.3}). While a resistance mutation emerges, the tumor grows faster under selection (despite the very low impact of therapy on sensitive cancer cells) and is therefore more likely to be detected at earlier times. By the time of detection, non-resistant tumors usually accumulate up to 4 additional drivers on average, while resistant tumors have fewer. For larger values of \(c\), an additional non-regularity emerges (segment \(BCD\) in \fref{3.1.4}), appearing at  \(\sigma\approx3s\) and is associated with tumors having a majority of cells with 3 total drivers. This region is also characterized by a different transition to complete resistance (compare Suppl. Video S1 and S2 for relatively low and high costs of resistance, respectively). For example, at point \(B\) tumors with a majority of non-resistance have less variable detection times than tumors with a majority of resistant cells (\textbf{A} and corresponding panel \textbf{B} in \fref{3.1.4}). Treatment levels along the segment \(BCD\) result in tumors that are more likely to be resistant as one goes from the center to the tails of the distribution \(H_M(t)\). This differs qualitatively from the previous case of low cost of resistance, where the tumors are less resistant in a tail of the distribution.

\begin{figure}[!tbh]\centering{\includegraphics[scale=0.56]{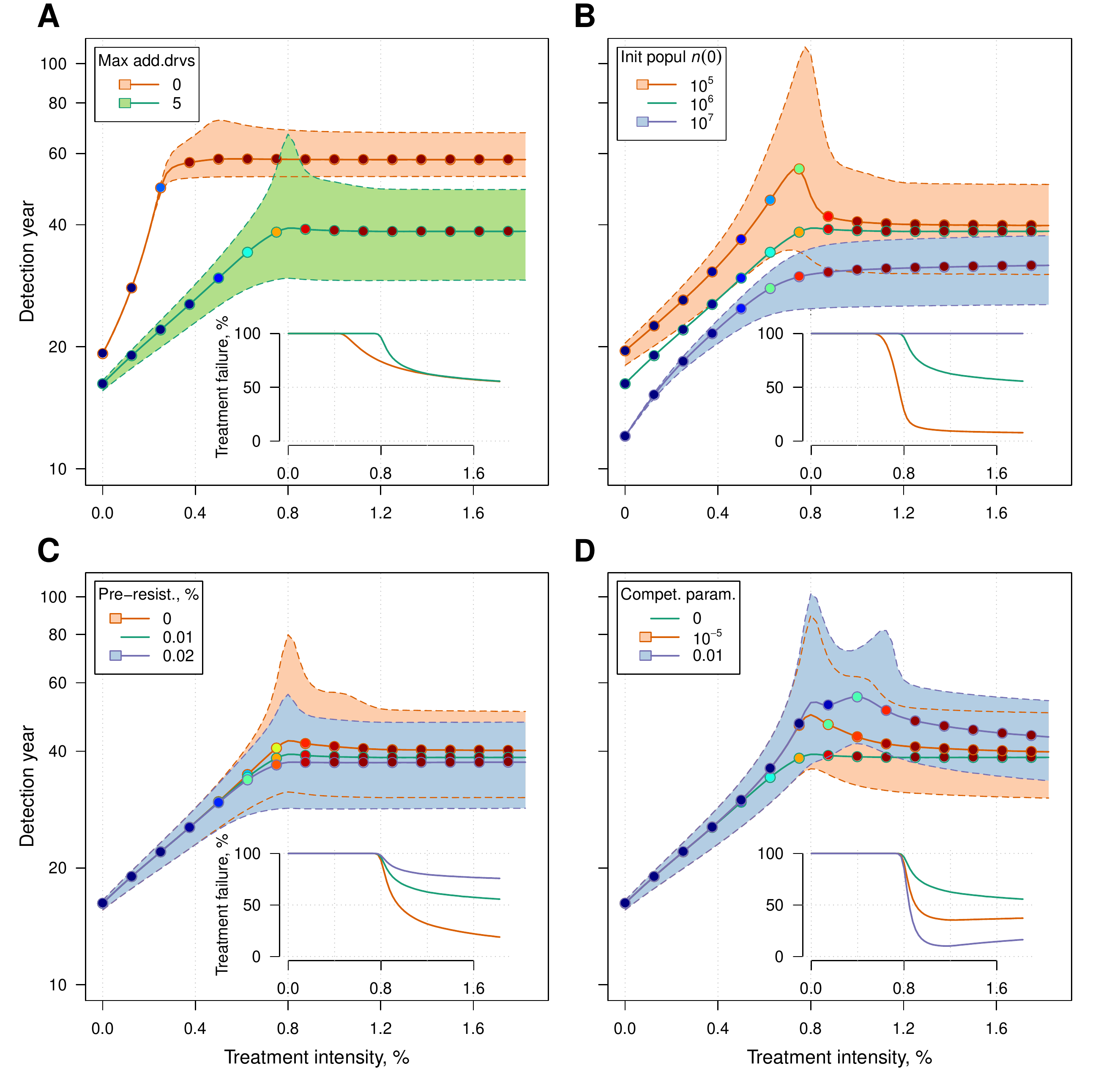}\quad}\caption{\textbf{Sensitivity analysis of several key parameters.} The median (think line) and 95\% confidence intervals (shaded areas with dashed boundaries) for the distribution of detection times. Parameter values are as in \tref{1}, except the one being varied: (\textbf{A}) maximal number of additionally accumulated drivers; (\textbf{B}) initial cell number \(n(0)\); (\textbf{C}) level of initial partial resistance of a tumor; (\textbf{D}) competitive parameter value \(\alpha\). The competition is implemented by introducing an exponential factor in the fitness calculation: \(f_{i0} = s(i+1)-\sigma\) and \(f_{i1} = s(i+1)e^{-\alpha S}-c\), where \(S=\sum_{i}n_{i0}\) is the number of non-resistant cells and \(\alpha\) characterizes the strength of competition. The insets show failure with respect to the change in the treatment level. The color-code for points indicates the average level of resistance within tumors, analogous to \fref{3.1.3}. The \textbf{insets} show the percentage of cases in simulations, leading to detection of a tumor rather than its extinction or keeping its size below the detection threshold. For simplicity of representation, only the median is indicated in \textbf{B}, \textbf{C} and \textbf{D} for the baseline case, which is shown in green in \textbf{A} and with all parameter values, indicated in \tref{1}.}\label{fig:3.1.9}\end{figure}

The initial cell number \(n(0)\) affects both the median and distribution of detection times \fref{3.1.9}(\textbf{B}). For large initial tumors, growth is deterministic and exponential. As the initial size is decreased (\(n(0)=10^6\) to \(10^5\)), stochastic effects are increasingly manifested by the appearance of an inflection point in the trajectory of the median, as well as increased variability in detection times.

Whereas \(M=10^9\) cells are an approximate clinical detection threshold for many solid tumors, approximately \(\bar M=10^{11}\) cells could be attained for certain otherwise undetectable cancers, and are only discovered in late stage, metastatic disease \cite{AktiBodd13}. Using numerical experiments, we find that the difference in times when a tumor reaches \(M\) and \(\bar M\) respectively, is \(4.7\pm0.2\) years (mean\(\pm\)s.d.).

A tumor is likely to be eradicated under a range of constant treatments if it has \(n(0)=10^5\) or fewer initial cells; a tumor is virtually certain to persist for \(n(0)=10^7\) cells or greater, as it is shown in \sfref{3.1.6}. In other words, our model indicates that tumors that are c. 1\% of clinically detectable size will typically be impossible to eradicate. 

The above analysis assumes zero initial resistance within a tumor. Given mutation rates assumed here we can expect that many tumors with one million cells will already contain resistant cells. We extend our study to other values of initial resistance level, denoted by \(\kappa\). As shown in \fref{3.1.9}(\textbf{C}) larger values of \(\kappa\) create a transition from stochastic to deterministic tumor growth. As expected, larger \(\kappa\) results in worse control outcomes, with a threshold for treatment failure - tumors can only be eradicated for \(\sigma>2s\) (see the inset \fref{3.1.9}(\textbf{C})).

Finally, we briefly consider how competition affects the results presented above. We assume for simplicity that sensitive cells inhibit the growth of resistant cells (i.e., chemoresistance has both a constant cost and an additional cost in proportion to sensitive cell number). Such assumption in its simple form leads to more variability in tumor growth and, as expected, delays in cancer detection \fref{3.1.9}(\textbf{D}) and a positive correlation with treatment success, see the inset \fref{3.1.9}(\textbf{D}), which can be understood as the proportion of numerical experiments in which the tumor stays undetected. \section{Discussion} 

Maximum tolerated dose chemotherapies present numerous challenges, ostensibly the major one being the selection of resistant phenotypes, which are possible precursors for relapse \cite{GerlSwan10}. Over the past decade, several alternative approaches have been proposed, where the objective is to manage rather than eradicate tumors (e.g., \cite{KomaWoda05,Gate09,GateSilv09,GateBrow09,MaleReid04,FooMich10}. Tumor management attempts to limit cancer growth, metastasis, and reduce the probability of obtaining resistance mutations through micro-environmental modification or through competition with non-resistant cancer cell populations or with healthy cells. These approaches usually involve clinically diagnosed cancers: either inoperable tumors or residual cancers after tumor excision. In the former situation tumors are typically large enough in size to contain numerous resistance mutations. In many, if not most, cases these neoplasms will have metastasized, meaning greater variability both in terms of phenotypes and hence potential resistance to chemotherapies, and in penetrance of therapeutic molecules to targeted tumor cells \cite{KleiBlan02,ByrnBouc05}. The latter situation involves smaller, residual cancer cell populations, but composed of high frequencies of resistant variants or dormant cells \cite{KleiBlan02}. Both scenarios are likely to involve cancer cell populations with large numbers of accumulated driver mutations, which ostensibly contribute to the speed of relapse. Thus, management of clinically detected tumors need not only limit the proliferation and spread of refractory subpopulations (\fref{3.1.3}-\ref{fig:3.1.9}), but should also aim to control the growth of multi-driver clones (\sfref{3.1.S2}).
 
We mathematically investigated an alternative strategy, chemoprevention through satisficing, where a satisfactory objective is defined from the outset for patients at a high risk of contracting a life-threatening cancer. Such objectives can be complex, involving minimal side effects, defining acceptable risks of developing a lethal cancer at a later time, and realistic maximum expected frequencies of chemoresistant lineages. Neoplasms in our model system could correspond to pre-cancerous states of dyplasia, carcinoma in situ, or to invasive carcinoma, but the relative frequency of these different stages for tumors of the initial sizes modeled here are unknown; nor is it known how chemopreventive therapies affect cell populations in these different states. Several authors have previously argued for how constant or intermittent low toxicity therapies either before or after tumor discovery could be an alternative to maximum tolerated dose chemotherapies \cite{WuXLipp11,HochThom13}, but to our knowledge no study has actually quantified the modalities (treatment start time, dose) for such approaches using empirically derived parameter estimates \cite{BeerAnta07,BoziAnta10,BoziReit13}. 

Our model indicates that daily reductions in population growth of \(s<\sigma<2s\), corresponding to \(0.1\%\)-\(0.2\%\), is sufficient in most cases to control neoplasm expansion for tumors less than about 1 million cells, and harboring no resistance mutations and only one driver mutation at the start of therapy. We find that such nascent tumors can be managed for tens of years without growing to life-threatening levels, and that the duration of successful prevention is sensitive to both initial tumor size and treatment intensity. Specifically, tumors growing beyond approximately the reciprocal of the driver mutation rate (c \(10^{6}\) cells) are exponentially increasingly likely to produce a faster growing subclone with a new driver. Such a subclone is even more likely, once it reaches \({\sim}1/u\) cells, to produce a new subclone with an additional driver, and so on. The result over sufficiently long periods is a hyper-exponential increase in tumor size. Given that mutation rates for chemoresistance are thought to be on the order of \(10^{-6}\) per cell division \cite{KomaWoda05}, this means that tumor size is also a sensitive predictor of the likelihood of chemoresistance, and thus the potential for chemopreventive management to slow the progression of a potentially lethal cancer. 

Indeed, one of our central results is the sensitivity of tumor growth to size at the commencement of therapy (\fref{3.1.9}(\textbf{C})). Deterministic equations provide an accurate description of such growth for sufficiently small (\(\lesssim10^5\)) or large (\({>}10^6\)) initial tumor sizes. We found that sufficiently small tumors were controlled and sometimes eradicated if therapeutic reductions in population growth exceeded \(2s\). Larger tumors, though affected by therapy, were impossible to control or to eradicate, because of the presence of resistance mutations. This effect was mitigated to some extent by cell-cell competition (\fref{3.1.9}(\textbf{D})), but the process of competition as modeled here was not sufficient to permit tumor control (but see \cite{GateSilv09}). Moreover, we found that tumors approximately between \(10^5\) and \(10^6\) cells had more variable, stochastic outcomes, meaning that a given preventive therapeutic regimen may or may not be successful due to the chance emergence of driver mutations and local extinctions. This result emphasizes not only the sensitivity of tumor control to initial size (i.e., time at which therapy commences), but also the accurate assessment of changes in cancer risk for different therapeutic alternatives. Although not investigated in the context of cancer therapies, the results of Bozic and colleagues \cite{BoziAnta10} indicate that achieving a tumor size where driver mutations become probable, distinguishes patients harboring small tumors after 25 years from those developing life-threatening tumors over this same period (see Fig. 1 in \cite{BoziAnta10}). In their simulations like ours, the time to emergence of the second driver mutation is a good predictor of future tumor growth. 

Our theory also highlights two potentially contrasting objectives of cancer prevention: managing tumor size vs managing resistance mutations. We show the sensitivity of these two outcomes to treatment levels, especially near the threshold \(2s\). Treatments at or beyond \(2s\) effectively offset or reduce sensitive cell growth, leaving a large subpopulation of resistant cells that are released from competition and can rapidly obtain additional driver mutations. Treatments just below \(2s\) also reduce tumor growth, but maintain high frequencies of sensitive cells, which potentially compete with resistance cells, thereby reducing overall resistance cell numbers. Adopting the latter strategy could make a difference to long-term outcomes, especially in cases where the constant therapy is discontinued, or ulterior attempts are made at high dose chemotherapy.

Some empirical studies support the role of certain molecules in chemoprevention \cite{WillHeym09}. For example, Silva and colleagues \cite{SilvKam12} parameterized computational models to show how low doses of verapamil and 2-deoxyglucose could be administered adaptively to promote longer tumor progression times. These drugs are thought to increase the costs of resistance and the competitive impacts of sensitive on resistant cancer cell subpopulations. However, some of the most promising results have come from studies employing non-steroidal anti-inflammatory drugs (NSAIDs), including experiments \cite{IbraCorn11}, investigations of their molecular effects \cite{KostKuhn13,GaliLiX07}, and their use \cite{CuziThor14}.  For example, Ibrahim and coworkers \cite{IbraCorn11} studied the action of NSAIDs and specifically sodium bicarbonate in reducing prostate tumors in male TRAMP mice (i.e. an animal model of transgenic adenocarcinoma of the mouse prostate). They showed that mice commencing the treatment at 4 weeks of age had significantly smaller tumor masses, and that more survived to the end of the experiment than either the controls or those mice commencing the treatment at an older age. Kostadinov and colleagues \cite{KostKuhn13} showed how NSAID use in a sample of people with Barrett's esophagus is associated with reductions in somatic genomic abnormalities and their growth to detectable levels. It is noteworthy that it is not known to what extent reductions in cancer progression under NSAIDs is due to either cytotoxic or cytostatic effects, or both. Although we do not explicitly model cytotoxic or cytostatic impacts, therapies curbing net growth rates, but maintaining them at or above zero, could be interpreted as resulting from the action of either cytotoxic and/or cytostatic processes. In contrast, therapies reducing net growth rates below zero necessarily have a cytostatic component. Our model, or modifications of it to explicitly include cytotoxic and cytostatic effects, could be used in future research to make predictions about optimal dose and start times to achieve acceptable levels of tumor control, or the probability of a given tumor size by a given age. Lorz and coworkers \cite{LorzLore13} recently modeled the employment of cytotoxic and cytostatic therapies alone or in combination and showed how combination strategies could be designed to be superior in terms of tumor eradication and managing resistance than either agent used alone. 

Decisions whether or not to employ chemopreventive therapies carry with them the risk of a poorer outcome than would have been the case had another available strategy, or no treatment at all, been adopted \cite{EsseSepu04}. This issue is relevant to all preventive approaches, where alterations in life-style, removal or treatment of pre-cancerous lesions, or medications may result in unwanted side effects or potentially induce new invasive neoplasms (e.g., \cite{BerrCurt11}). Chemopreventive management prior to clinical detection would be most appropriate for individuals with genetic predispositions, familial histories, elevated levels of specific biomarkers, or risk-associated behaviors or life-styles \cite{HochThom13,SutcHumm09,HemmLiX04,LippLee06,WillHeym09}. Importantly, our approach presupposes that the danger a nascent, growing tumor presents is proportional to its size and (implicitly, all else being equal) a person’s age. Due caution is necessary in applying our results, since studies have argued that metastatic potential rather than tumor size may be a better predictor of future survival \cite{FoulSmit10,Hyne03,SethKang11}. 

We have modeled preventive approaches to managing risks of future lethal cancers. However, our model also could be applied to scenarios where an invasive carcinoma is discovered early in progression. In such cases, tumor clones are likely to harbor greater numbers of driver mutations and show higher levels of genomic instability and standing genetic heterogeneity than the earlier stages targeted by chemoprevention. Higher aptitude for growth and mutation (adaptation) in clinical tumors could mean that outcomes are less sensitive to cancer cell numbers as we found in the prevention scenario. We suggest that the frequency distribution of driver mutations and the distribution of resistant subclones within these lineages could instruct decisions of the time course of treatment levels, with the aims of satisfactory tumor growth, metastasis, and resistance control. Although residual cancer cell populations from excised tumors and associated micro-metastases are often difficult to assess with accuracy \cite{PantCote99}, our results suggest that if order of magnitude estimates are possible, than low dose, constant approaches could be optimized, and according to our model, will always be superior to aggressive chemotherapies even if resistance mutations are likely to be present. \section{Models and methods} 
\subsection{Conceptual framework}

Let each cell in a population be described by two characteristics. The first is its resistance status, which is either ``not resistant'' (\(j = 0\)) or ``resistant'' (\(j = 1\)). The probability of a resistant mutant emerging during cell division is assumed constant (\(v\)). The second property is the number of accumulated driver mutations in a given cell line (maximum \(N\)). The mutation rate at any locus resulting in the addition of a driver is \(u\), and we assume no back mutation. Thus, the genome of a cell in our model is composed of \(N\) potential driver loci and one chemoresistance locus.

We initially assume that at each time step cells either divide or die, but do not compete for space or limiting resources. The fitness function \(f_{ij}\) is the difference between the birth and death rates of a cell and is defined by the number of accumulated drivers (\(i=0,1,\ldots,N\)) and resistance status (\(j=0,1\)). A chemosensitive cancerous cell with a single driver has selective advantage \(s\). Any additional drivers add \(s\) to fitness, while resistance is associated with a constant cost \(c\). Exposure to a single chemotherapy treatment affects only non-resistant cells (\(j=0\)), incurring a loss \(\sigma\in[0,1]\) to their fitness. We assume that all parameters \(c\), \(s\) and \(\sigma\) are arbitrarily small (\(\ll1\)).

Fitness is 
\begin{equation}\label{eqn:1.1} f_{ij} = s(i+1)-\sigma(1-j)-cj\,.    \end{equation}
The assumption of driver additivity is a special case of multiplicative fitness, and both are approximately equivalent for very small \(s\).

\subsection{Numerical simulations}

To simulate tumor growth, we adopt a discrete time branching process for the cell-division process \cite{BoziAnta10,Durr12}, which is usually referred to as a discrete time Galton-Watson process \cite{AthrNey72}. For each numerical experiment we initiate a tumor of a given size with cells of a type \(i=0\), and proportion of resistant cells within a tumor \(\kappa\) (\(0\le\kappa\le1\)). Table 1 presents baseline parameter values employed in this study.

At the beginning of each time step, the number of cells is \(n_{ij}(t)\). The number of cells at next step \((t+1)\) is then sampled by a multinomial distribution. If we let \(B_{ij}\) be the number of births in the population, \(D_{ij}\) the number of deaths, \(M^{(u)}_{ij}\) and \(M^{(v)}_{ij}\) the number of mutations from class \((i,j)\) to classes \((i+1,j)\) and \((i,j+1)\) respectively, then the multinomial distribution is 
\[ P[(B_{ij},D_{ij},M^{(u)}_{ij},M^{(v)}_{ij})=(k_1,k_2,k_3,k_4)]=\frac{n_{ij}(t)!}{k_1!k_2!k_3!k_4!}(b_{ij}(1-u_i-v_j))^{k_1}d_{ij}^{k_2}(b_{ij}u_i)^{k_3}(b_{ij}v_j)^{k_4}\,, \]
where \(u_i=u(1-i/N)\) and \(v_j=v(1-j)\) (\(i=0,\ldots,N\), \(j=0,1\)). 

The number of cells of type \((i,j)\) at time step \(t+1\) is now given by
\[ n_{ij}(t+1) = n_{ij}(t)+B_{ij}-D_{ij}+M^{(u)}_{i-1,j}+M^{(v)}_{i,j-1}\,, \]
where we assume \(M^{(u)}_{-1,j}=0\) and \(M^{(v)}_{i,-1}=0\).

We conducted numerical experiments of the above model, each with the same initial states, but each using a unique set of randomly generated numbers of a branching process. 

\subsection{Code}

All calculations were made using programs, written in C, and the free, open-source statistical package R \cite{Rpackage}. The color palette for figures was adopted from \cite{ColorBrewer2}. Code for all calculations, and for producing all of the figures, is available at \cite{SupplCode14} and can be used freely for non-commercial purposes.

\subsection{Mean-field dynamics}

We use the mean-field approach, see e.g. \cite{KrapRedn10}, which approximates the behavior of a system consisting of many cells, so that the effects of stochasticity are averaged and an intermediate state is described by a set of ordinary differential equations. 

\paragraph{Master equations}

We write master equations to track the probability \(P_{ij}(t)\) that a randomly chosen cell from the population of tumor cells will be of type \((i, j)\) at time \(t\). 

The temporal dynamics of probabilities \(P_{ij}(t)\), \(i = 0, 1, \ldots, N\), where \(N\) is the maximal number of additionally acquired drivers and \(j = 0, 1\), are described by
\[ \frac{\mathrm dP_{ij}(t)}{\mathrm dt}=\mathbb P_{ij}+u\mathbb P_{ij}^{(u)}+v\mathbb P_{ij}^{(v)}\,. \]
Here, the right-hand side is a superposition of probabilistic in- and out-flows from different mutational states to the current one \((i, j)\). The function \(\mathbb P_{ij}\) describes the growth of subclone \((i, j)\) and is proportional to the probability \(P_{ij}(t)\), multiplied by the difference between \(f_{ij}\) and the average fitness over the whole population  \(\bar f=\sum_{i,j}f_{ij}P_{ij}(t)\). \(\mathbb P_{ij}^{(u)}\) and \(\mathbb P_{ij}^{(v)}\) represent the probabilistic flows of mutations. For \(\mathbb P_{ij}^{(u)}\), a driver is added from class \((i-1, j)\) to \((i, j)\) in proportion to the probability \(P_{i-1,j}(t)\), the probability of cell birth \(b_{i-1,j}\), and the probability of a zero locus being chosen from \(N\) total loci consisting of \(N - (i - 1)\) other zero loci. A similar approach is used to define the outflow term for the probability from class \((i, j)\) to \((i + 1, j)\). The second term \(\mathbb P_{ij}^{(v)}\) is the probability of mutating to therapeutic resistance (\((i, j = 0)\) to \((i, j = 1)\)), and is proportional to \(P_{i0}(t)\) and birth rate \(b_{i0}\). Finally, all terms are summed, taking into account the initial conditions: \(P_{00}(0) = 1-\kappa\), \(P_{01}(0) = \kappa\) and \(P_{ij}(0) = 0\) for any other \(i\) or \(j\).

The above elements lead to the following system of ordinary differential equations (ODEs):
\begin{equation}\label{eqn:2.1.1} \begin{split} \frac{\mathrm dP_{ij}(t)}{\mathrm dt}=(f_{ij}-\bar f)P_{ij}(t)+u\left[\bigl(1-\frac{i-1}N\bigr)\frac{1+f_{i-1,j}}2P_{i-1,j}(t)-\bigl(1-\frac iN\bigr)\frac{1+f_{ij}}2P_{ij}(t)\right]-{}\\v(1-2j)\frac{1+f_{i0}}2P_{i0}(t)\,, \end{split}    \end{equation}
where some probabilities \(P_{ij}\) could, theoretically, take on negative values, e.g. \(P_{-1,j}(t)\), when \(i=0\), in which case, they are set to zero. 

A simple transformation
\[ p_{ij}(0)=P_{ij}(0),\quad p_{ij}(t)=P_{ij}(t)\exp\bigl(\int_0^t\bar f(r)\>\mathrm dr\bigr)\,, \]
allows omitting the term \(\bar f\) from equation \eref{2.1.1} and to linearize the latter with respect to the new ``transformed'' probabilities \(p_{ij}(t)\). This gives
\begin{equation}\label{eqn:2.1.2} \begin{split} \frac{\mathrm dp_{ij}(t)}{\mathrm dt}=f_{ij}p_{ij}(t)+u\left[\bigl(1-\frac{i-1}N\bigr)\frac{1+f_{i-1,j}}2p_{i-1,j}(t)-\bigl(1-\frac iN\bigr)\frac{1+f_{ij}}2p_{ij}(t)\right]+{}\\v\frac{1+f_{ij}}2(jp_{i,j-1}(t)+(j-1)p_{ij}(t))\,, \end{split}    \end{equation}
where, for convenience, we write \((jp_{i,j-1}(t)+(j-1)p_{ij}(t))\) instead of \((1-2j)p_{i0}(t)\).

\paragraph{Probability generating function approach}

With the master equations \eref{2.1.2}, we apply the probability generating function (p.g.f.) method \cite{Gard04,Assa10} to transform the system of \((2N + 1)\) ordinary differential equations to a  Hamilton-Jacobi (HJ) equation, that is, a first order partial differential equation. 

We define the p.g.f. as the polynomial over all modified probabilities \(p_{ij}\) of the form
\begin{equation}\label{eqn:2.2.1} G(\xi,\eta,t)=\sum\limits_{i=0}^N\sum\limits_{j=0}^1 \xi^i\eta^jp_{ij}(t)\,,    \end{equation}
where \(\xi\) and \(\eta\) are variables that can be viewed as the momentum of an auxiliary Hamiltonian system governing the leading-order stochastic dynamics of the system \cite{ElgaKame04}. Notice that the function \(G(\xi,\eta,t)\) is linear with respect to \(\eta\).

Suppose that the function \(G(\xi,\eta,t)\) is defined, one can then obtain all characteristics of the stochastic process such as the average tumor size \(n(t)\) and the average frequency \(n_{res}(t)/n(t)\) of resistant cells within a tumor. The former quantity is 
\[ \frac{\mathrm dn(t)}{\mathrm dt}=n(t)\bar f(t)\,. \]
Using the normalization condition for the probability: \(\sum_{i,j}P_{ij}(t)=1\), we obtain
\[ G(\xi=1,\eta=1,t)=\exp\bigl(\int_0^t \bar f(r)\>\mathrm dr\bigr)\,, \]
and then
\begin{equation}\label{eqn:2.2.2} n(t)=n(0)\exp\bigl(\int_0^t\bar f(r)\>\mathrm dr\bigr)=n(0)G(\xi=1,\eta=1,t)\,,    \end{equation}
where the initial tumor size \(n(0)\) is sufficiently large. The latter quantity is written as follows
\begin{equation}\label{eqn:2.2.3} \frac{n_{res}(t)}{n(t)}=\sum\limits_{i=0}^NP_{i1}(t)=\sum\limits_{i=0}^Np_{i1}(t)\exp\bigl({-}\int_0^t\bar f(r)\>\mathrm dr\bigr)=\frac{\partial G/\partial\eta}{G(\xi,\eta,t)}\Big|_{\xi=1,\eta=1}\,.    \end{equation}

Initial conditions yield \(p_{00}(0)=1-\kappa\), \(p_{01}(0)=\kappa\) and \(p_{ij}(0)=0\) for any other \(i\) or \(j\), so that \(G(\xi,\eta,t=0)=1-\kappa+\kappa\eta\).

To obtain the HJ equation related to the p.g.f. \(G(\xi,\eta,t)\), we multiply \eref{2.1.2} on \(\xi^i\eta^j\) and sum up all equations for \(i=0,1,\ldots,N\) and \(j=0,1\). After some algebra, we obtain
\begin{equation}\label{eqn:2.2.4} \frac{\partial G}{\partial t}=\left[s\bigl(\xi\frac{\partial}{\partial\xi}+1\bigr)-\sigma\bigl(1-\eta\frac{\partial}{\partial\eta}\bigr)-c\eta\frac{\partial}{\partial\eta}+\frac{u(\xi-1)}2\bigl(1-\frac\xi{N}\frac{\partial}{\partial\xi}\bigr)+\frac{v(\eta-1)}2\bigl(1-\eta\frac{\partial}{\partial\eta}\bigr)\right]G\,,    \end{equation}
where only terms of order greater than or equal to \(u\), \(v\) are retained, meaning that terms composed of the products \(s\), \(c\) and \(u\), \(v\) are omitted.

Equation \eref{2.2.4} is solved by the method of characteristics such that the HJ equation is transformed into a system of ordinary differential equations (i.e., the system of characteristics, see e.g. \cite{Meli98}).

\paragraph{Constant treatment}

We study the case for constant \(\sigma\). Notice that this includes the case of no treatment (\(\sigma=0\)).

First, we find the characteristics for the variables \(\xi\) and \(\eta\). Namely, we write using \eref{2.2.4}:
\begin{equation}\label{eqn:2.3.1} \frac{\mathrm d\xi(t)}{\mathrm dt} = -s\xi(t)+\frac{u\xi(\xi-1)}{2N}\,, \frac{\mathrm d\eta(t)}{\mathrm dt}=(c-\sigma)\eta+\frac{v\eta(\eta-1)}{2}\,,    \end{equation}
which gives
\begin{equation}\label{eqn:2.3.2} \xi(t) = \frac{s+u/(2N)}{Ae^{(s+u/(2N))t}+u/(2N)}\,,\quad \eta(t) = \frac{\sigma-c+v/2}{Be^{(\sigma-c+v/2)t}+v/2}\,,    \end{equation}
where \(A\) and \(B\) are integration constants associated with initial values of \(\xi(0)\) and \(\eta(0)\) as 
\begin{equation}\label{eqn:2.3.3} \xi(0) = \frac{s+u/(2N)}{A+u/(2N)}\,,\quad \eta(0) = \frac{\sigma-c+v/2}{B+v/2}\,.    \end{equation}

The p.g.f. \(G(\xi,\eta,t)\) changes along the characteristic \eref{2.3.2}-\eref{2.3.3} according to the following ODE
\[ \frac{\mathrm dG(t)}{\mathrm dt} = \bigl(s-\sigma+\frac{u(\xi(t)-1)}2+\frac{v(\eta(t)-1)}2\bigr)G(t)\,, \]
which is straightforward to integrate. Indeed, if we use \eref{2.3.1}, this yields: \(\mathrm d\ln G = (s(N+1)-c)\mathrm dt + N\mathrm d\ln\xi + \mathrm d\ln\eta\). Then, with initial condition \(G(\xi(0),\eta(0),0)=(1-\kappa)+\kappa\eta(0)\), \(\kappa\) is a level of resistance within a tumor (\(\kappa\in[0,1]\)), \eref{2.3.2} and \eref{2.3.3}, we finally obtain the solution to \eref{2.2.4} of the following form
\[ \begin{split} G(\xi,\eta,t) = G(\xi(0),\eta(0),0)\exp\left[(s-\sigma-(u+v)/2)t + N\ln\bigl(1+\frac{\xi u}{2N}\frac{e^{(s+u/(2N))t}-1}{s+u/(2N)}\bigr) + {} \right. \\ \left.\ln\bigl(1+\frac{\eta v}{2}\frac{e^{(\sigma-c+v/2)t}-1}{\sigma-c+v/2}\bigr)\right]\,. \end{split} \]
Taking into account \(u,v\ll s,c\) and assuming \(v\ll \sigma-c\), we can simplify \eref{2.3.4} further and write its approximate form
\[ G(\xi,\eta,t) \approx \Big(1-\kappa+\frac{\kappa\eta e^{(\sigma-c)t}}{1+\frac{\eta v}2\frac{e^{(\sigma-c)t}-1}{\sigma-c}}\Big)\exp\left[(s-\sigma)t + N\ln\bigl(1+\frac{\xi u}{2N}\frac{e^{st}-1}{s}\bigr) + \ln\bigl(1+\frac{\eta v}{2}\frac{e^{(\sigma-c)t}-1}{\sigma-c}\bigr)\right]\,, \]
which can be further simplified and written in the form
\begin{equation}\label{eqn:2.3.4} G(\xi,\eta,t) \approx \left((1-\kappa)(1+\frac{\eta v}{2}\frac{e^{(\sigma-c)t}-1}{\sigma-c})+\kappa\eta e^{(\sigma-c)t}\right)\exp\left[(s-\sigma)t + N\ln\bigl(1+\frac{\xi u}{2N}\frac{e^{st}-1}{s}\bigr)\right]\,.    \end{equation}
As expected \eref{2.3.4} is linear with respect to \(\eta\).

The dynamics for the frequency of resistant cells within a tumor \eref{2.2.3} is then given by 
\begin{equation}\label{eqn:2.3.5} \frac{\partial G}{\partial\eta} \approx \Big((1-\kappa)\frac{v}{2}\frac{e^{(\sigma-c)t}-1}{\sigma-c}+\kappa e^{(\sigma-c)t}\Big)\exp\!\left[(s-\sigma)t + N\ln\bigl(1+\frac{\xi u}{2N}\frac{e^{st}-1}{s}\bigr)\right]\,.   \end{equation} \section*{Acknowledgement}

The authors are grateful to Athena Aktipis, Sylvain Gandon, Urszula Hibner, Patrice Lassus and Carlo Maley for discussions and helpful remarks. ARA thanks all members of Evolutionary Community Ecology group (University of Montpellier 2), especially Marie Vasse, Sarah Calba and Isabelle Gounand, for their support. This work was made possible by the facilities of the Shared Hierarchical Academic Research Computing Network (SHARCNET:www.sharcnet.ca) and Compute/Calcul Canada.
\bibliographystyle{unsrtnat}
\bibliography{biblio}

\setcounter{figure}{0}
\setcounter{equation}{0}
\makeatletter 
\renewcommand{\thefigure}{S\@arabic\c@figure}
\renewcommand{\theequation}{S\@arabic\c@equation}
\makeatother

 \section*{Supplemantary material} 
\begin{figure}[!tbh]\centering{\includegraphics[scale=0.59]{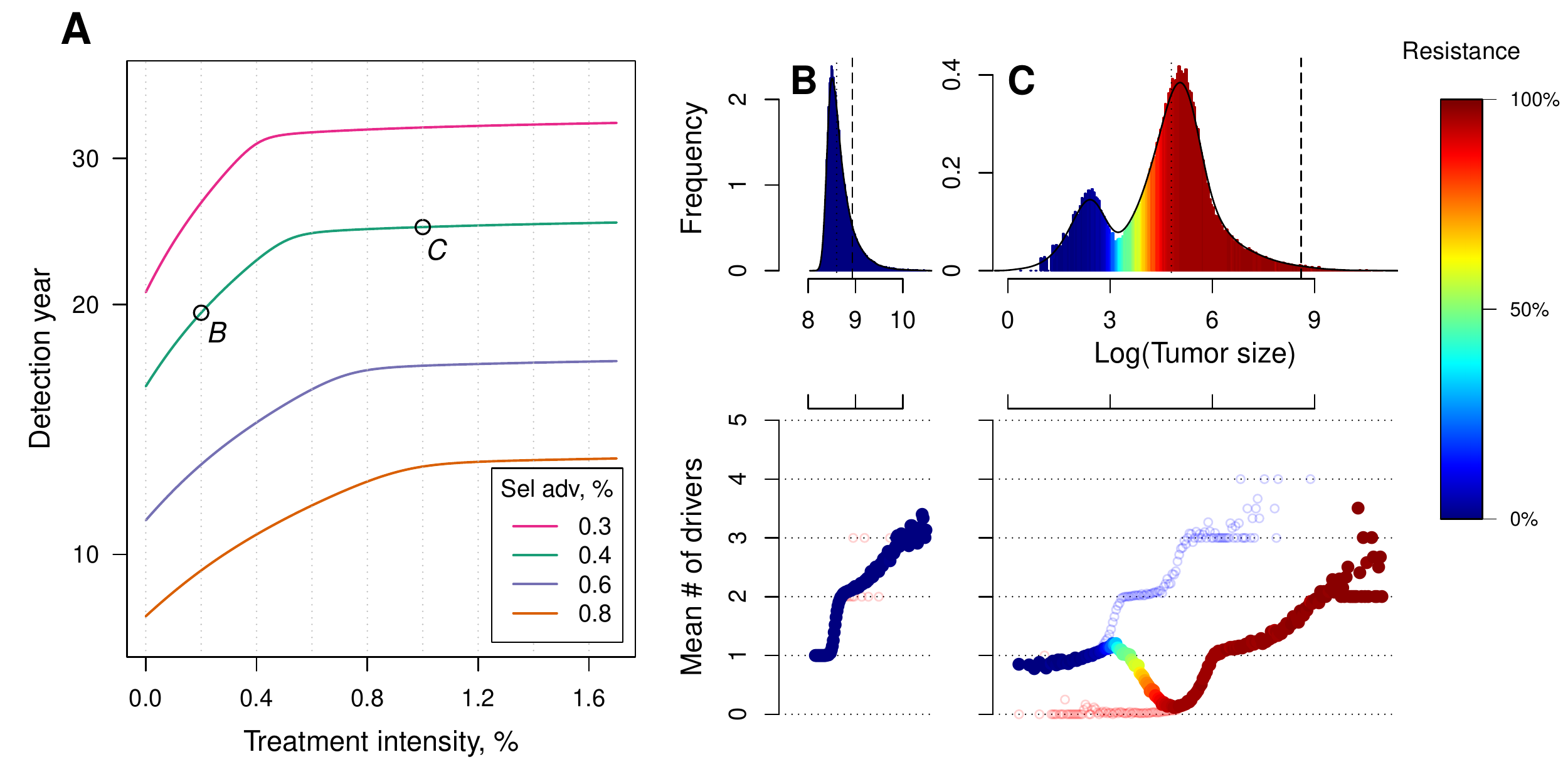}\quad}\caption{\textbf{Tradeoff between growth and resistance under different treatment regimes.} (\textbf{A}) Analytically-derived times for a tumor to reach \(10^9\) cells (see equation \eref{3.1.1}). (\textbf{B} and \textbf{C}) Sample distributions for corresponding points \emph{B} and \emph{C}, shown in plot \textbf{A}. The bottom panel shows the mean number of additionally accumulated drivers for all detected tumors over intervals of 3 months. Light red points indicate tumors with a majority of resistant cells, while light blue points are for tumors dominated by non-resistant cells. The color-code indicates the level of resistance in detected tumors over 3 month intervals (see the colorbar on the right for details). Parameters otherwise as in \tref{1}. }\label{fig:3.1.S1}\end{figure}

\begin{figure}[!tbh]\centering{\includegraphics[scale=0.59]{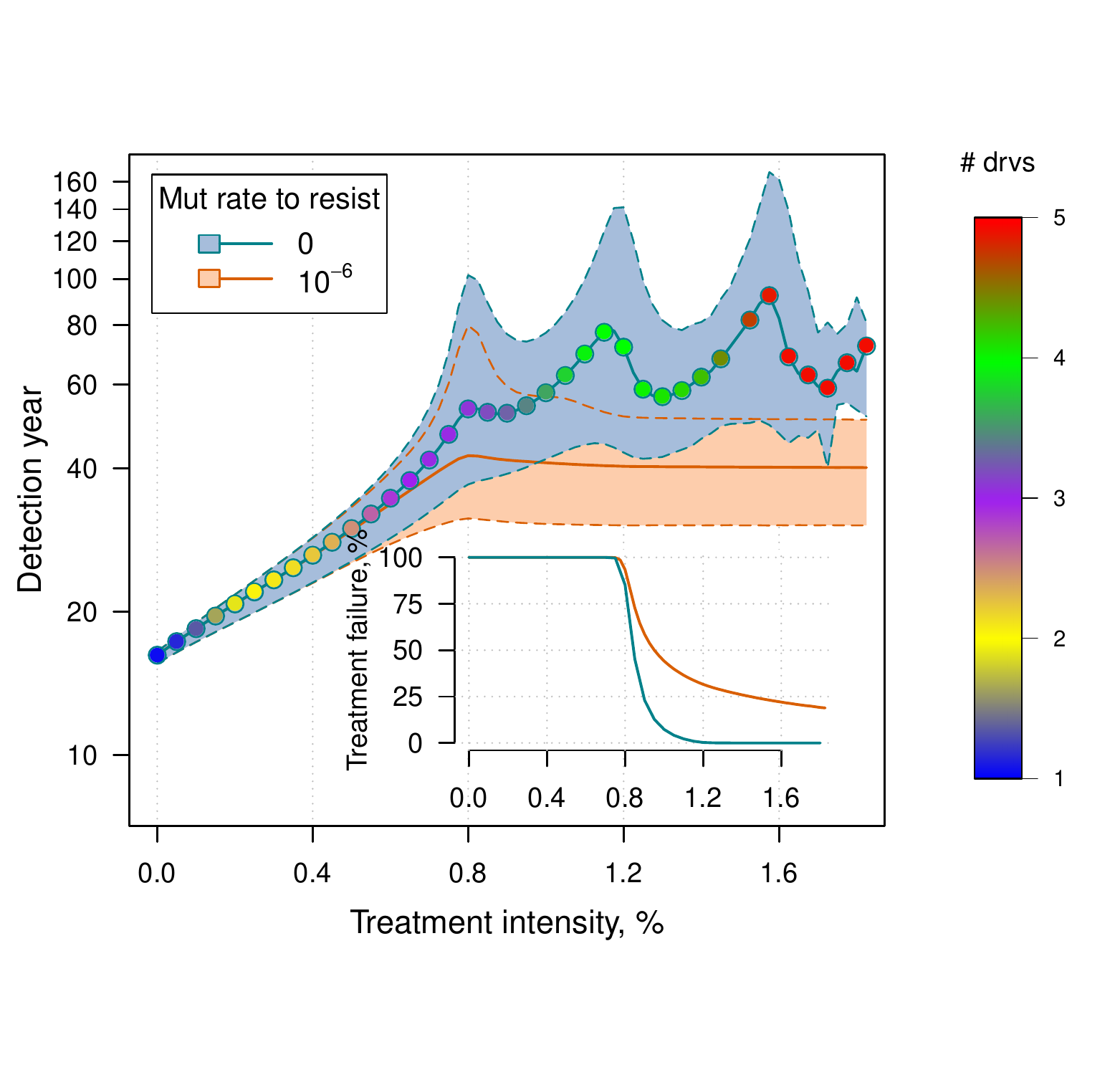}\quad}\caption{\textbf{In the absence of resistance, higher treatment selects for faster growing subclones.} The median (thick blue) and 95\% confidence intervals (shaded blue area) for the distributions of detection times, when a resistant mutation is knocked out. (no initial level of resistance of a tumor (\(\kappa=0\)) and zero mutation rate to acquire the resistance \(v=0\)). For comparizon, the case of non-zero mutation rate \(v\) for the same initial conditions is shown besides in orange. The point color-code indicates the average number of additionally accumulated drivers within detected tumors. }\label{fig:3.1.S2}\end{figure}

\begin{figure}[!tbh]\centering{\includegraphics[scale=0.6]{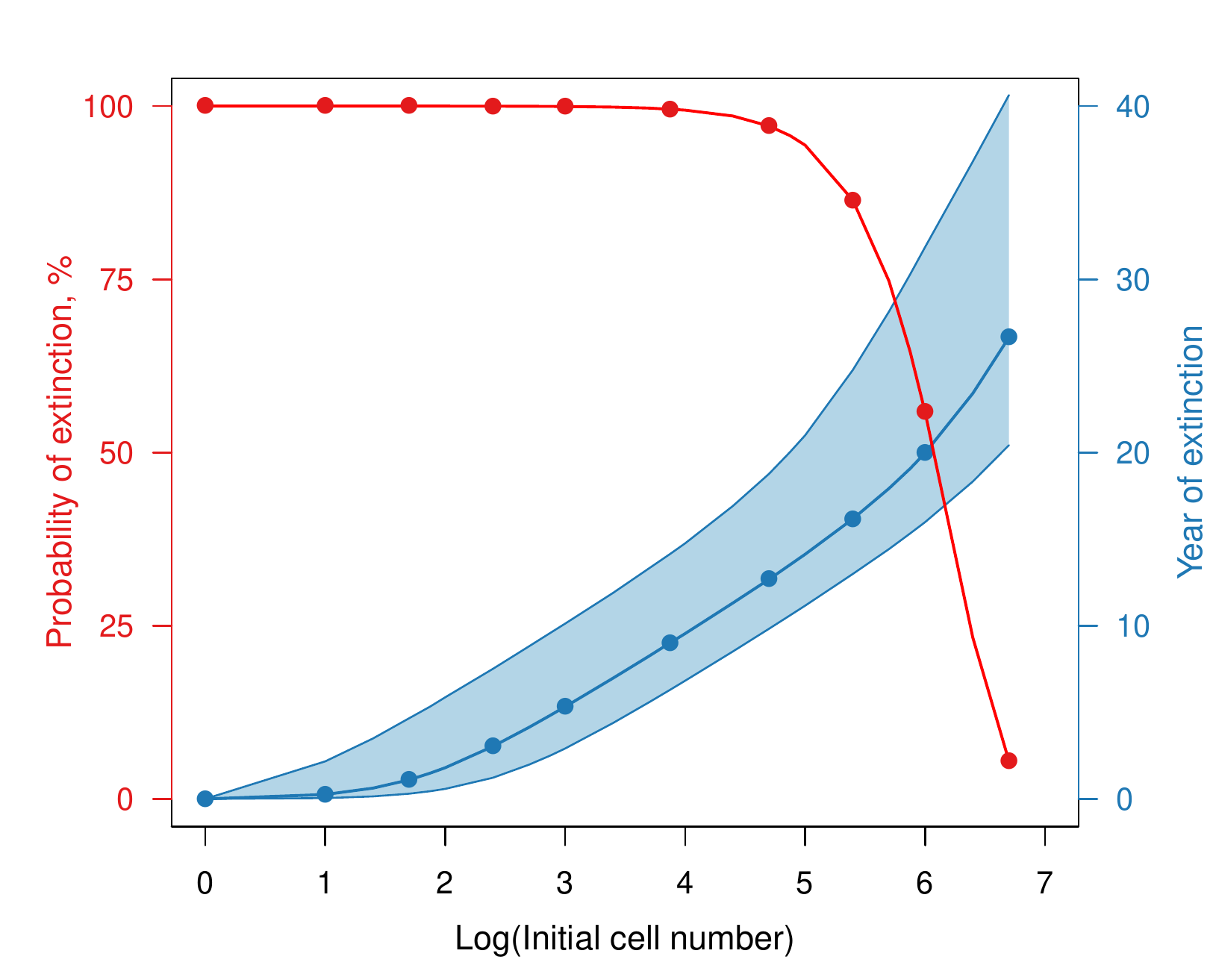}\quad}\caption{\textbf{Sufficiently small tumors can be driven to extinction by low dose therapies.} The median (thick blue) and 95\% confidence intervals (shaded area) for the distributions of extinction times. Red line indicates the probability of extinction, depending on initial cell number. Treatment level is  \(0.1\%\), and no pre-resistance \(\kappa=0.0\). Parameters otherwise as in \tref{1}.}\label{fig:3.1.6}\end{figure}
\end{document}

%% file: table1.tex
\begin{center}
	\begin{tabular}[tbh]{l|c|c|c}
		\bf Parameter & \bf Variable & \bf Value & \bf Reference \\ \hline
		Time step (cell cycle length) & $T$ & 4 days & \cite{BoziAnta10} \\ 
		Selective advantage & $s$ & $0.4\%$ & \cite{BoziAnta10} \\ 
		Cost of resistance & $c$ & $0.1\%$ & \\ 
		Mutation rate to acquire an additional driver & $u$ & $3.4\times10^{-5}$ & \cite{BoziAnta10} \\ 
		Mutation rate to acquire resistance  & $v$ & $10^{-6}$ & \cite{KomaWoda05} \\ 
		Maximal number of additional drivers & $N$ & 5 & \\ 
		Initial cell population & $n(0)$ & $10^6$ & \\ 
		Pre-resistance level & $\kappa$ & $0.01\%$ & \cite{IwasNowa06} \\
		Number of replicate numerical simulations & - & $10^6$ & \\ 
		(excl. the ones with extinction) & & & 
	\end{tabular}
\end{center}